\documentclass[amssymb,aps,prli,preprint]{revtex4-1}
\usepackage{latexsym,amsmath,graphics,graphicx,exscale,colordvi,color,mathtools}
 \pdfoutput=1
\usepackage{soul,xcolor,braket}

\setstcolor{red}

\begin{document}

\title{Unoccupied surface and interface states in Pd thin films deposited on Fe/Ir(111) surface }
\author{Mohammed Bouhassoune}\email{m.bouhassoune@fz-juelich.de}
\author{Imara L. Fernandes}
\author{Stefan Bl\"ugel}
\author{Samir Lounis}\email{s.lounis@fz-juelich.de}
\affiliation{Peter Gr\"unberg Institut \& Institute for Advanced Simulation, 
Forschungszentrum J\"ulich \& JARA, D-52425 J\"ulich,
Germany}

\begin{abstract}
{We present a systematic first-principles study of the electronic surface states and resonances occuring in thin films of Pd of various thicknesses deposited on a single ferromagnetic monolayer of Fe on top of Ir(111) substrate. This system is of interest since one Pd layer deposited on Fe/Ir(111) hosts small magnetic skyrmions. The latter are topological magnetic objects with swirling spin-textures with possible implications in the context of spintronic devices since they have the potential to be used as magnetic bits for information technology. The stabilization, detection and manipulation of such non-collinear magnetic entities require a quantitative investigation and a fundamental understanding of their electronic structure. Here we investigate the nature of the unoccupied electronic states in Pd/Fe/Ir(111), which are essential in the large spin-mixing magnetoresistance (XMR) signature captured using non spin-polarized scanning tunnelling microscopy [Crum et al., Nat. Commun. {\bf 6} 8541 (2015); Hanneken et al., Nat. Nanotech. {\bf 10}, 1039 (2015)]. To provide a complete analysis, we investigate bare Fe/Ir(111) and Pd$_{n=2,7}$/Fe/Ir(111) surfaces. Our results demonstrate the emergence of surface and interface states after deposition of Pd monolayers, which are strongly impacted by the large spin-orbit coupling of Ir surface.}

\end{abstract}
\maketitle


\section{Introduction}
Understanding the surface electronic structure is one of the key ingredients for designing and developing new functionalities in information technology. Recently, it was demonstrated that the electronic states residing on a surface hosting non-collinear magnetic states can give rise to a large magnetoresistance effect, which was coined the spin-mixing magnetoresistance (XMR) effect~\cite{Crum} or non-collinear magnetoresistance (NCMR)~\cite{Hanneken}. The current flow between the substrate and a non-magnetic electrode was found to depend strongly on the spin-mixing induced by the misalignment of the magnetic moments and on 
the presence of spin-orbit interaction in the probed  magnetic electrode. This effect is thus  different from the giant magnetoresistance (GMR)~\cite{Grunberg, Fert1} or tunneling magnetoresistance (TMR) effects~\cite{Julliere}, requesting two magnetic electrodes during measurements.

After discovering in Pd/Fe/Ir(111)\cite{Romming} magnetic skyrmions, which are swirling spin-textures of topological nature~\cite{Bogdanov,Nagaosa}, this system has been heavily studied~\cite{Crum,Hanneken,Romming15,Dupe,Simon,Santos,Imara,Kubetzka,Boettcher,Malottki,Bessarab}, however not much is known about its electronic properties. Obviously, the surface electronic properties, which are important for the XMR effect, are shaped by the layer facing vacuum, \textit{i.e.} Pd, after state hybridization involving Fe and the underlying Ir interface layers. 
 
Various studies have been performed on films made of either pure Fe, Pd or Ir layers (see \textit{e.g.}~\cite{Turner,Turner2,Ohnishi,Brooks,Braun,Bisi,Louie,Hulbert,Corso,Busse,Pletikosic}). Since Ir is a heavy element, spin-orbit interaction (SOI) can play an important role. For example, it leads in fcc-Ir(111) substrate to a giant Rashba splitting of an occupied surface state that is rather robust to surface coverage with graphene~\cite{Varykhalov} as demonstrated with  angle-resolved photoemission and \textit{ab-initio} simulations.  
 
While the surface state of clean fcc-Ir(111) substrate is occupied~\cite{Corso}, it is unoccupied in fcc-Pd(111)~\cite{Hulbert}. This aspect is interesting since Ir and Pd are neighbors in the periodic table. 
  In fact, if one exchanges Ir with another neighboring element from the periodic table being valence-isoelectronic to Pd, e.g. Pt, the unoccupied surface state is restored at 0.5 eV above the Fermi energy~\cite{Roos,Wiebe}. Because of the weaker ionic potential of Pd compared to that of Pt, the unoccupied surface state of the former is located at a higher energy (1.3 eV)~\cite{Hulbert}. 
  The system Pd/Fe/Ir(111) provides the playground to investigate how the interplay of interface and confinement effects with magnetism, inducing a spin-dependent shift of the potential, affects the aforementioned  surface electronic features. 
  
By depositing Pd on top of the ferromagnetic Fe grown on Ir(111) surface and by focusing our first-principles study on the unoccupied states related to the observed large XMR effect, we find interface states confined within PdFeIr interfaces besides an unoccupied surface state of Pd. The latter is demonstrated by observing its stability upon the increase of the thickness of Pd films. We analyze the impact of SOI and find that it can project into a given spin-channel, confined states originally living in the opposite spin-channel.
 
The article is structured as follows. First, we describe the methodology and the computational details followed for the simulations then present our results and discussions. Previous experimental studies have shown the possibility of growing Pd in either hcp- or fcc-stacking on top of fcc-Fe/Ir(111) surface~\cite{Hanneken,Kubetzka}. As shown in Refs.~\cite{Dupe}, there are differences in the electronic structure between these two stacking leading to a shift of the unoccupied electronic features. To make our study concise, we focus here on the fcc-stacking of Pd as a follow up of our previous study~\cite{Crum}. We address initially the two-dimensional Bloch spectral function of fcc-Pd/Fe/Ir(111) surface and discuss the nature of the different unoccupied states showing high intensity at $\overline\Gamma$, the center of the two-dimensional Brillouin zone.  By investigating the case of bare fcc-Fe/Ir(111) surface, which was shown to host a lattice of skyrmions in its ground state~\cite{Heinze,Dupe2,Hauptmann}, and considering thicker Pd films deposited on that surface, we aim at distinguishing the states stemming from Pd.
Our conclusions are summarized in the last section.

\section{Method}
We use the full-potential Korringa-Kohn-Rostocker Green function method~\cite{Bauer} including SOI as implemented within density functional theory. The exchange and correlation effects are treated in the local spin-density approximation as parametrized by Vosko, Wilk and Nusair~\cite{Vosko}. 
We consider slabs of 44 layers to model the ferromagnetic Pd$_{n{\mathrm{MLs}}}$/Fe/Ir(111), with $n=1,2,...,7$ being the number of monolayers (MLs). PdFe layers are deposited on one side of the slab in contrast to the vacuum layers being on each side. The vacuum layers are represented by empty spheres shown as white balls in Fig. 1. 
We incorporated the layer positions found from the simulations reported by Dup\'e et al.~\cite{Dupe} and assume an fcc-stacking of the layers (see Figure 2). The self-consistent calculations are performed using an angular momentum cutoff of $l$ = 3 for the orbital expansions of the Green functions and a grid of $30 \times 30$ k-points mesh for the sampling of the two-dimensional Brillouin zone. The complex-energy contour needed for the energy integration is rectangular and consists of 40 grid points including seven Matsubara frequencies. For the calculation of the local density of states (LDOS), the Brillouin zone is sampled with a $200 \times 200$ k-points mesh. The surface-projected band structure is calculated along the following high symmetry directions $\overline\Gamma \overline K$, $\overline K \overline M$ and $\overline M \overline\Gamma$.    
Furthermore, to have an adequate interpretation and analysis of our results we have considered a semi-infinite Ir (111) substrate using the decimation technique~\cite{Moliner,Szunyogh} to avoid any size effect that can blind our conclusions. Also, we have performed single-shot calculations without spin-orbit coupling in order to assess its effect on the k-dependent density of states. In the discussion focusing on the unoccupied states, all energies shown are positive and given with respect to the Fermi energy.

\begin{figure}
\begin{center}
\includegraphics*[angle=0,width=1\linewidth]{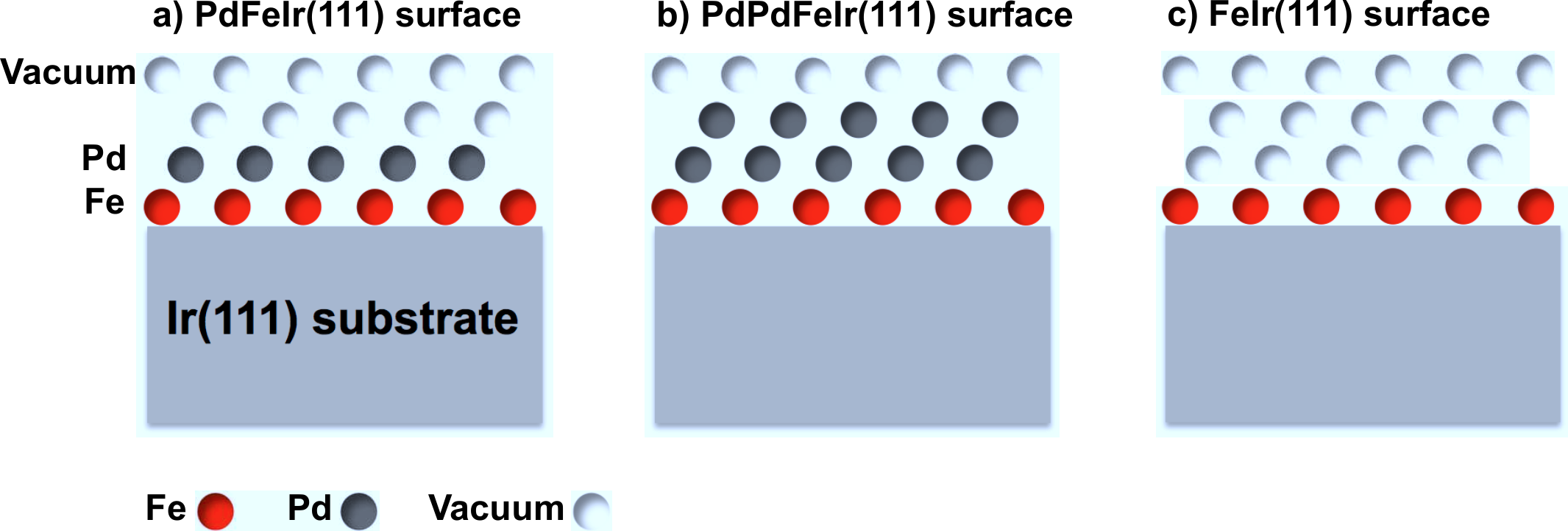}
\end{center}
\caption{Illustration of the hetero-structure of a) fcc-Pd/Fe/Ir(111), b) fcc-Pd$_{\mathrm{2}}$/Fe/Ir(111) and c) bare  Fe/Ir(111).}  
\label{2A1fig1}
\end{figure}

\section{Results and Discussion}
\subsection{Unoccupied states of Pd/Fe/Ir(111) and Fe/Ir(111) surfaces}
\begin{figure}[ht!]
\begin{center}
\includegraphics*[angle=0,width=0.48\linewidth]{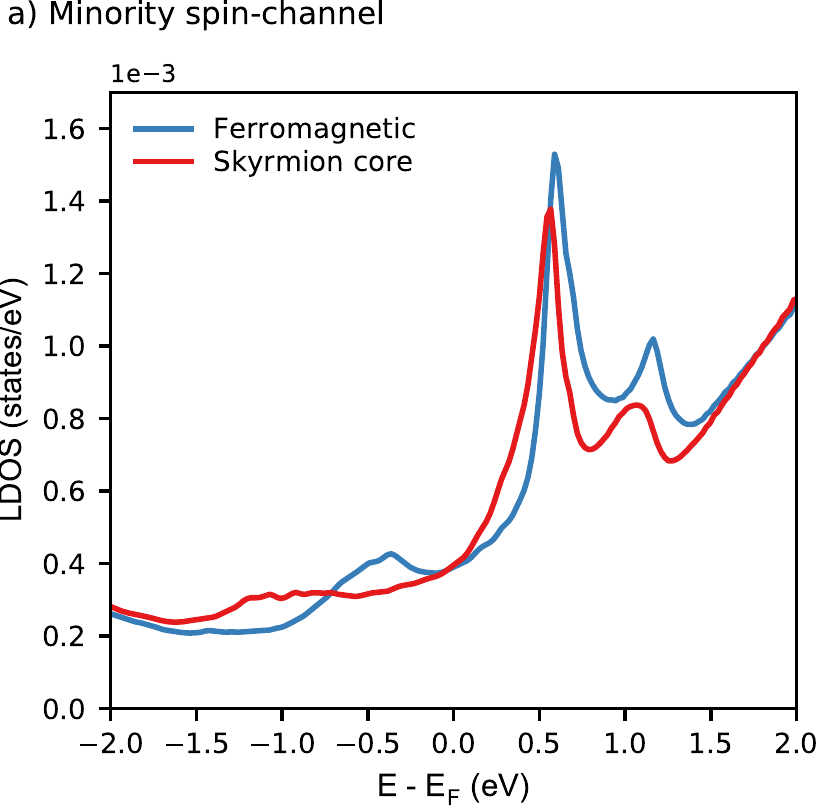}
\hspace{0.1cm}
\hfill
\includegraphics*[angle=0,width=0.48\linewidth]{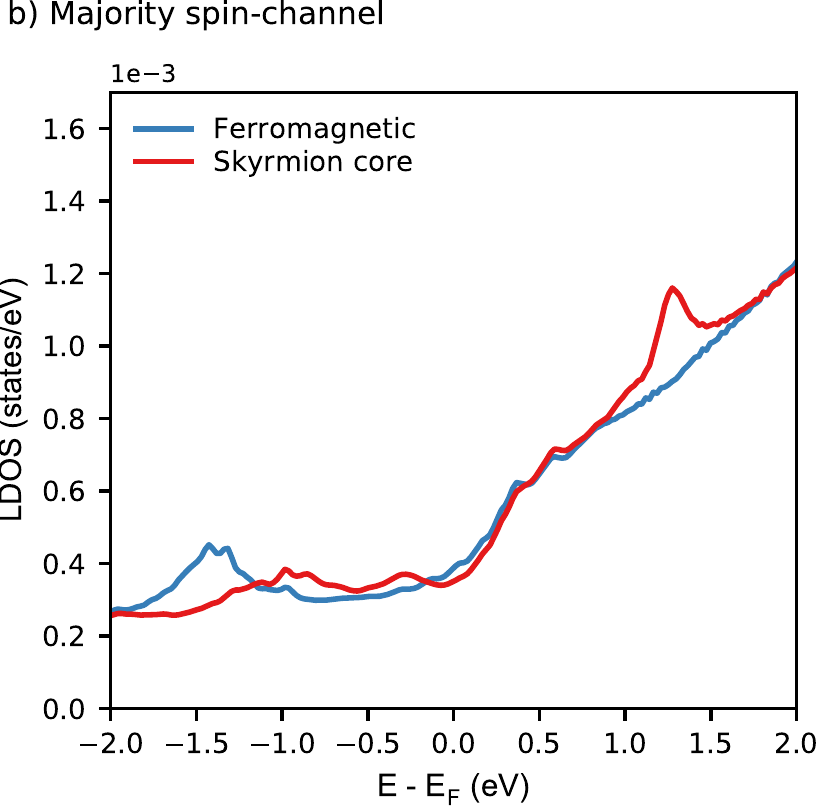}
\end{center}
\caption{ The local density of states (LDOS) of the vacuum site atop the core of the magnetic skyrmion (red) compared to that of the ferromagnetic state in Pd/Fe/Ir(111) surface as obtained from previous simulations~\cite{Crum}. The minority and majority spin-channels are depicted respectively in (a) and (b).}
\label{1A1fig1}
\end{figure}

To motivate our interest in studying the unoccupied states of Pd/Fe/Ir(111) surface, we present in Fig.~\ref{1A1fig1} the  local density of states (LDOS) obtained in previous ab-initio simulations~\cite{Crum}. The LDOS is calculated in vacuum at a distance of 2.2 {\AA} above the PdFe/Ir(111) surface in a ferromagnetic state (blue line) or above the core of a 17{\AA} wide skyrmion (red line). Here the magnetic skyrmion is simulated via an embedding technique based on the full-potential Korringa-Kohn-Rostoker Green function method with spin-orbit coupling included self-consistently.

The LDOS in vacuum can be related to the differential conductance measured with scanning tunneling microscopy (STM) according to the Tersoff-Hamann model~\cite{Tersoff} since STM probes the electronic states of the substrate surviving in vacuum at the tip position. One notices that the LDOS depends on the magnetic nature of the substrate. Our focus here is on the unoccupied states located at 0.68 eV, 1.27 eV and 1.45 eV above the Fermi energy in the minority-spin channel. In the majority-spin channel, one notices a step-like behavior at 0.37 eV and a feature around 1.3 eV on top of the skyrmion core. The latter feature is generated after spin-mixing from the states observed in the minority spin-channel of the ferromagnetic configuration (see Ref.~\cite{Crum} for extensive discussion). Note that once spin-orbit coupling is included, spin is not a good quantum number, thus minority- and majority-spin type of electrons are given in the local spin-frame of references pertaining to their respective sites. 

Figure 3 illustrates the surface-projected band structure of the Pd/Fe/Ir(111) system along $\overline\Gamma \overline K$, $\overline K \overline M$ and $\overline M \overline\Gamma$ directions for the minority- and majority-spin channels. Here spin-orbit interaction is included and the focus is on the unoccupied states. The dispersion curves are plotted for Ir ML (Figs. 3(a) and (e)) at the interface, Fe ML (Figs. 3(b) and (f)), Pd ML (Figs. 3(c) and (g)) and the first vacuum ML (Figs. 3(d) and (h)) on top of the surface. 

We are interested in the vacuum layer since according to Tersoff-Hamann~\cite{Tersoff} model, and as stated above, STM probes the electronic states of the substrate surviving in vacuum at the tip position. This allows then to understand the nature of the states triggering the large tunneling XMR shown recently~\cite{Crum,Hanneken}. 
Interestingly, besides several band gaps of the surface-projected band structure shown in Fig. 3 there are various high-intensity features living in both the minority- and majority-spin channels along the high-symmetry lines. The intensity of the states is obviously layer dependent. For instance, we notice a high intensity feature located around 1 eV at the $\overline M$ point of the majority-spin channel of Ir ML at the interface with Fe ML, which strongly decays into vacuum. Also, most of the intense minority-spin states of the Fe ML  do not survive in the vacuum layer.

The decay of a state into vacuum is k-dependent and approximately proportional to $e^{-\sqrt{\frac{2m\epsilon}{\hbar^2} + k}}$ with $\epsilon$ being the state's energy given with respect to the height of the potential barrier~\cite{Heinze2}. Therefore, it is expected that the states contributing most to the STM spectra are close to the center of the Brillouin zone, which motivates the focus of our following discussion on the surface-projected band structure at the ${\overline\Gamma}$ point. Indeed, one notices that away from ${\overline\Gamma}$, the electronic states experience a strong decay into the vacuum.\\

\begin{figure}
\begin{center}
\includegraphics*[angle=0,width=1\linewidth]{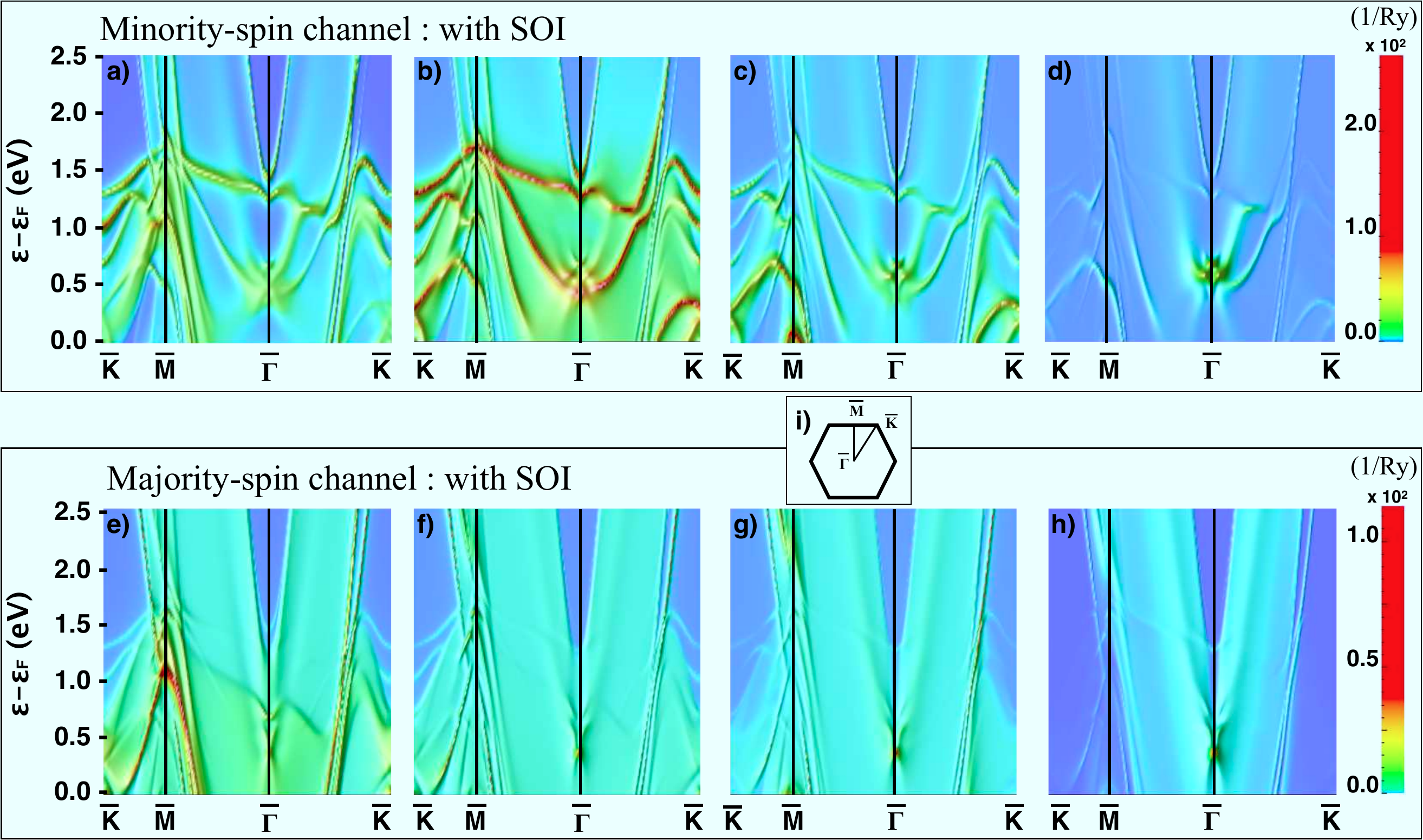}
\end{center}
\caption{Layer-resolved surface-projected band structure of Pd/Fe/Ir(111) including spin-orbit coupling for the minority- (a, b, c, d) and majority-spin (e, f, g, h) channels  along high-symmetry directions of the two-dimensional Brillouin zone (i): (a, e) Ir interface layer, (b, f) Fe interface layer, (c, g) Pd surface layer and (d, f) vacuum layer. We note that the reference energy is given by the Fermi energy $\epsilon_F$.}
\label{1fig1}
\end{figure}

\noindent{\bf Minority-spin channel of Pd/Fe/Ir(111).} Figs. 3 (a--d) show several bands with high-intensity crossing the $\overline\Gamma$ point, originating from different kinds of hybridization mechanisms.
For the Fe ML (Fig. 3 (b)), the most intense unoccupied minority-spin bands are located around 0.5 eV, being the broadest one, at 0.68, 1.27 and at 1.45 eV. These states are observed in the other MLs shown in Fig. 3 but with their relative intensity that can change. 
Note that the exact locations are obtained from the k-dependent DOS at the $\overline\Gamma$ point. The state at 1.45 eV can be either a surface or an interface state since it is located at the edge of the surface-projected  bulk-band gap similarly to the unoccupied surface state of Pt(111) substrate~\cite{Wiebe,Roos}. This contrasts with the other intense states, which are of resonant nature since they live in a``sea" of surface-projected bulk-bands. 

Our analysis indicates that the mentioned high-intensity states are located mainly at the IrFe and FePd interfaces, which are obviously hybrid states involving Ir, Fe and Pd layers. These states contribute to the STM spectra as one can notice that their intensities are high in vacuum with the largest contribution coming from the resonant states near 0.5 eV. This can be explained by their orbital nature and symmetry. 
$\Delta_1$ symmetry combining $s$, $p_z$ and $d_z^2$ 
substrate's orbitals provide the best matching to the vacuum's symmetry, which is invariant to rotations normal to the surface~\cite{Mavropoulos}.

\begin{figure}[ht!]
\begin{center}
\includegraphics*[angle=0,width=0.48\linewidth]{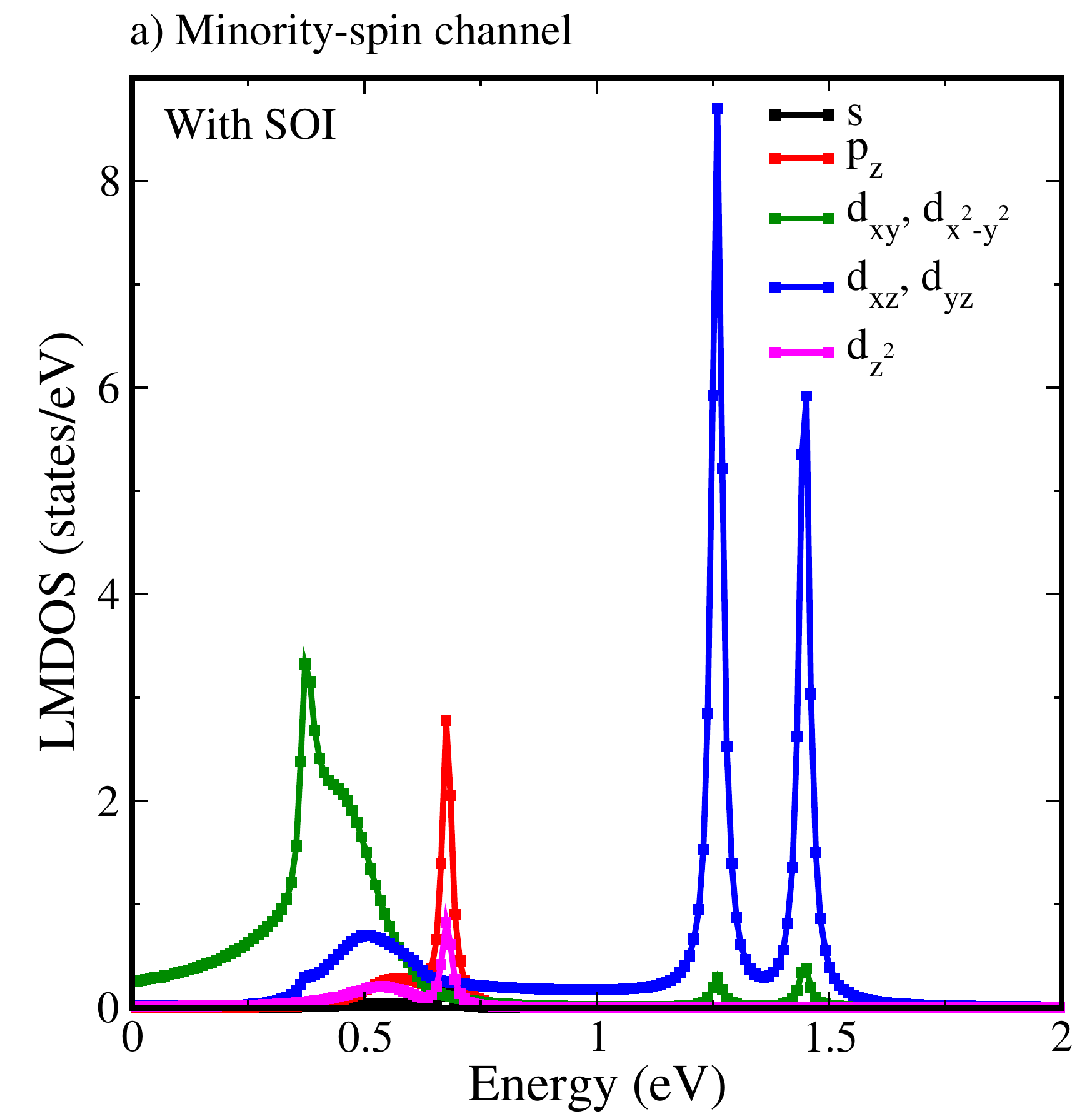}
\hspace{0.2cm}
\hfill
\includegraphics*[angle=0,width=0.48\linewidth]{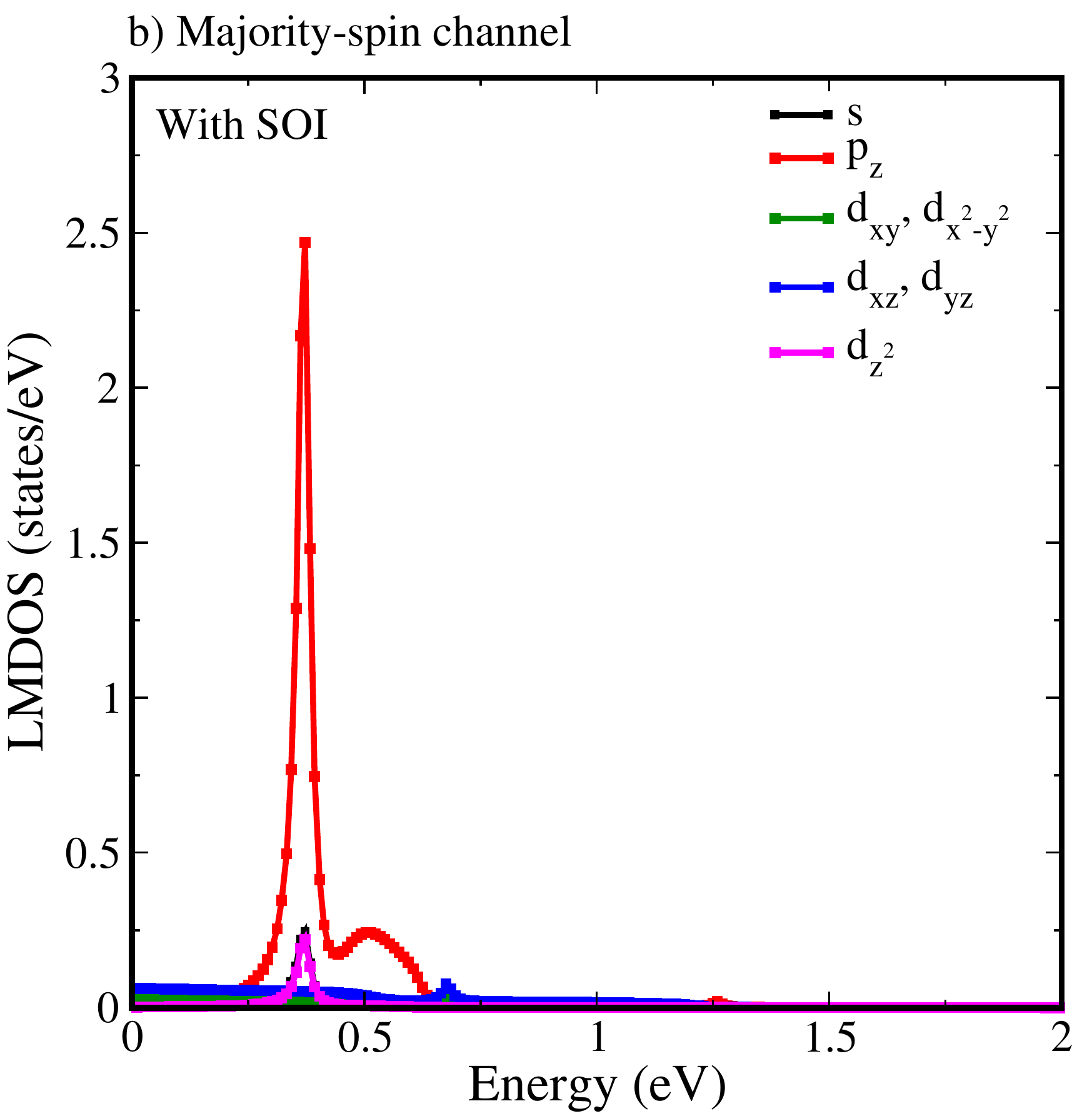}
\end{center}
\caption{Fe-orbital-resolved density of states (LMDOS) at the ${\overline\Gamma}$ point in Pd/Fe/Ir(111) surface for the (a) minority- and (b) majority-spin channels. Note that energies are given with respect to the Fermi energy.} 
\label{3fig3}
\end{figure}

\begin{figure}[ht!]
\begin{center}
\includegraphics*[angle=0,width=0.48\linewidth]{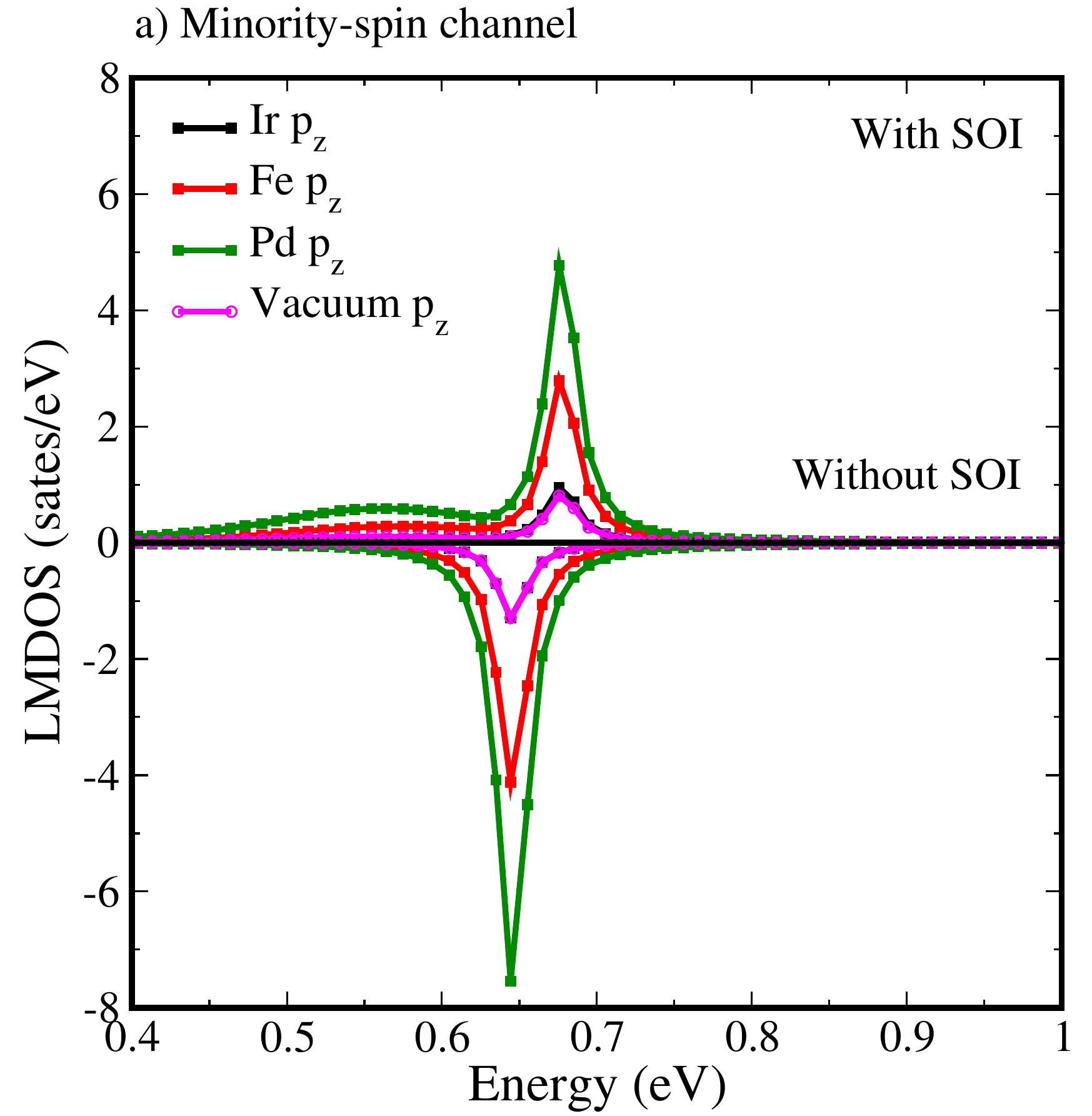}
\hspace{0.2cm}
\hfill
\includegraphics*[angle=0,width=0.48\linewidth]{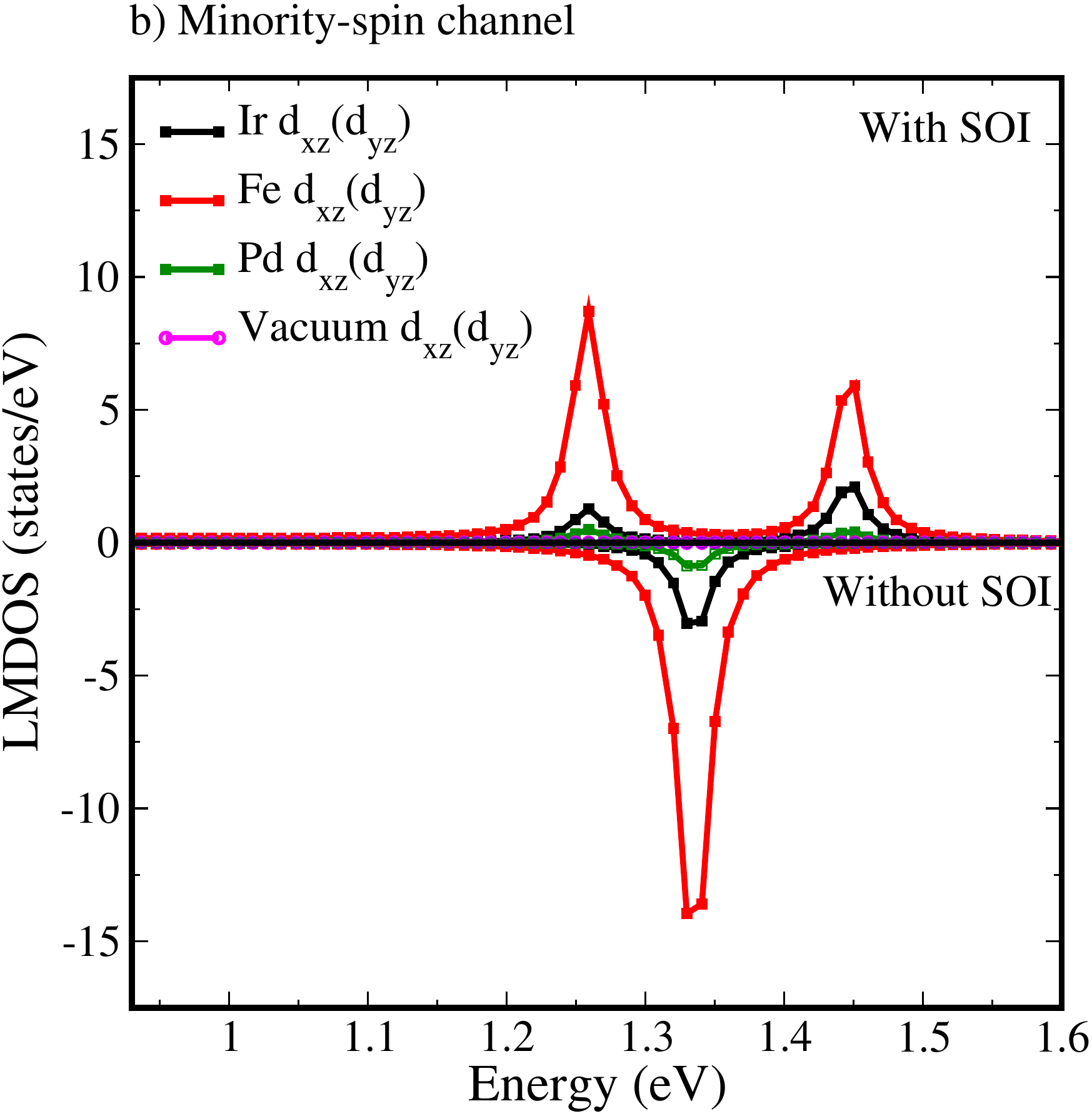}
\end{center}
\caption{Orbital-resolved density of states (LMDOS) at the ${\overline\Gamma}$ point for the minority-spin channel of Pd, Fe, Ir and vacuum layers: a) $p_{z}$ orbitals, b) $d_{xz}$ and $d_{yz}$ orbitals. In both figures the insets show both states for a large energy window (from 0 to 2.5 eV). Note that energies are given with respect to Fermi energy.} 
\label{4fig4}
\end{figure}
In order to unveil the origin of these states, we provide in Fig. 4 the orbital-resolved DOS (LMDOS) of the Fe ML at the $\overline\Gamma$-point. By analysing the localization of the different features in the slab, we came to the conclusion that in the minority-spin channel, the states at 1.27 and 1.45 eV are hybrid states of Ir and Fe and are mainly located in the Fe ML with respectively 80\% and 70\% of their total amplitude in IrFePd trilayer as can be seen in Fig. 5-b. Here, the trilayer consists of the Ir interface, Fe and Pd monolayers. 
These states are mainly $d_{xz}$ and $d_{yz}$ and thus are expected to decay much stronger than the $p_z$ or $d_{z^2}$ located in the lowest resonant state at 0.68 eV. (see Figs. 4-a and 5-a). The latter states stem mainly from the Fe $d_{z^2}$ and $p_z$ states hybridizing with the Pd states (with 50\% localisation in Pd, 38\% in Fe and 12\% in Ir of the total amplitude of these states in IrFePd trilayer). 
The hybridized bands appearing near these particular states, \textit{i.e.} at 0.5 eV, are a large mixture of Ir, Fe and Pd states mostly localized in Fe ML, $\approx$ 80\% of the sum of maximum amplitudes in the IrFePd trilayer, followed by Ir ($\approx$ 14\%) and finally Pd ($\approx$ 6\%).  
These bands are thus strongly localized in the Fe ML and characterized by a strong decay into the vacuum. When comparing to the sum of the amplitudes of the three other states discussed so far, \textit{i.e.} located at 0.68, 1.37 and 1.45 eV, the hybridized bands at 0.5 eV contribute by approximately only $14\%$ of the total amplitude collected in the vacuum layer. Once more, this is explained by their orbital nature, which dictates the tunneling amplitude.\\

\noindent{\bf Minority-spin channel of Fe/Ir(111).} Our statement on the origin of the observed states is confirmed after analyzing the surface-projected band structure of pure Fe/Ir(111) surface (Fig. 6 (a)). The states stemming from the hybridization with Pd electronic states around 0.68 eV and 0.5 eV in Pd/Fe/Ir(111) are very much affected in their shape and location after the removal of the Pd overlayer. For instance, we observed in bare Fe/Ir two largely splitted bands: one located slightly above 0.5 eV and the other around 0.2 eV.\\

\begin{figure}
\begin{center}
\includegraphics*[angle=0,width=1\linewidth]{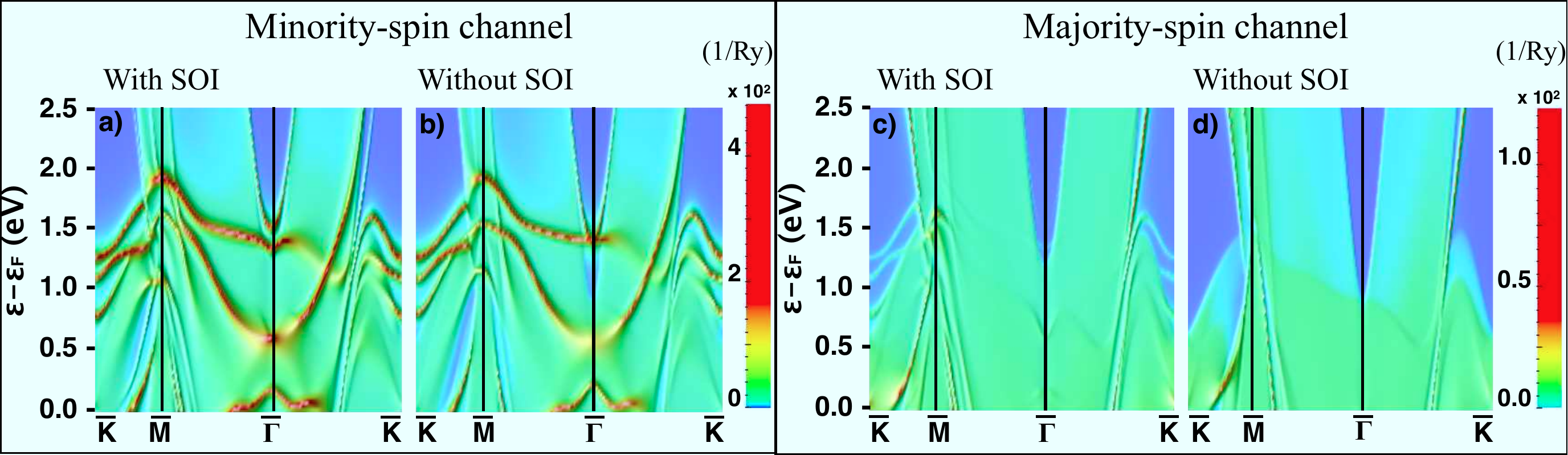}
\end{center}
\caption{Fe-ML's contribution to the surface-projected band structure of bare  Fe/Ir(111) with SOI (a,c) and without SOI (b,d). The minority-spin channel is plotted in (a) and (b) while the majority-spin channel is shown in (c) and (d).}
\label{1A1fig6}
\end{figure}

\noindent{\bf Impact of SOI on minority-spin channel.} It is interesting to consider the impact of SOI by switching it off and generating the surface-projected band structure starting from an already converged potential. The resulting band structure for Pd/Fe/Ir(111) is illustrated in Fig. 7, where one notices three bands crossing the ${\overline{\Gamma}}$ point, a degenerate band at 1.34 eV, a band at 0.64 eV and hybridized bands around 0.5 eV. 

\begin{figure}
\begin{center}
\includegraphics*[angle=0,width=\linewidth]{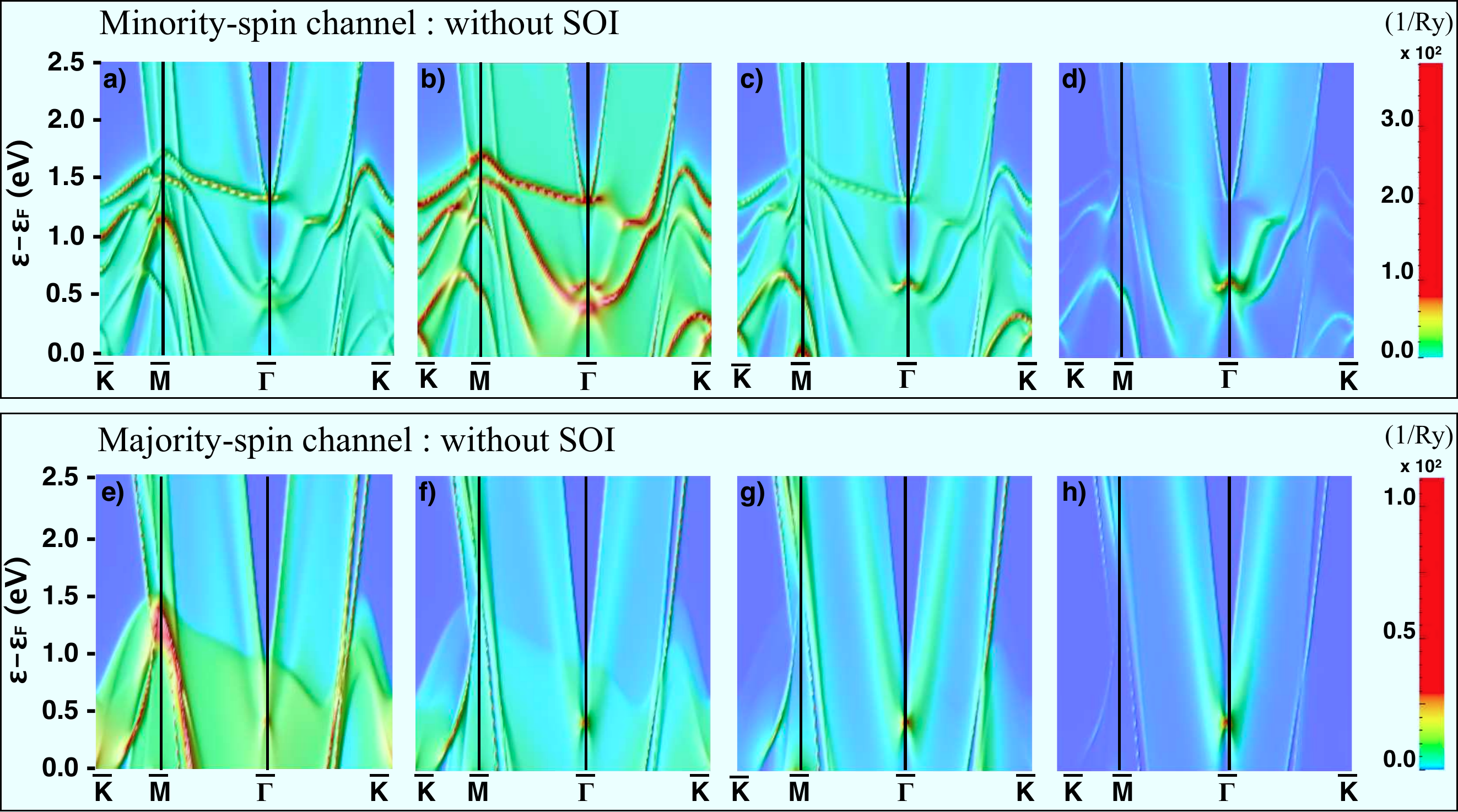}
\end{center}
\caption{Layer-resolved surface-projected band structure of Pd/Fe/Ir(111) without SOI for the minority-spin channel in: a) Ir interface layer, b) Fe Interface layer, c) Pd surface layer and d) vacuum layer. Following the same sequence layers, the surface-projected band structure for the majority-spin channel is plotted in (e, f, g, h). }
\label{1fig2}
\end{figure}

From these calculations, we can trace the origin of the resonant state located at 1.27 and of the interface state at 1.45 eV obtained when including SOI. Obviously, the initially degenerate bands at 1.34 eV get splitted by nearly 0.2 eV due to the large spin-orbit coupling introduced by the Ir substrate (see also Fig. 5(b)). As mentioned earlier, these states stem mainly from Ir and Fe hybrid states. Thus, SOI allows to promote an initially resonant state into an interface state at the edge of the gap of the surface-projected band structure. A similar behavior can be observed for bare Fe/Ir(111) surface by comparing Fig. 6(a) to Fig. 6(b). The band at 1.34 eV splits by approximately the same amount obtained for Pd/Fe/Ir(111). Also, the band at 0.68 eV in Pd/Fe/Ir(111) obtained when including SOI seems to be the result of the splitting of the band originally present at 0.64 eV without SOI (compare Fig. 3(b) with Fig. 6(b)). This can be better seen in Fig. 5(a). The splitting is about 0.1 eV, which indicates that the SO-driven splitting is state dependent. Note that in this region Pd states contribute a lot and since Pd is lighter than Ir, a lower impact of SOI is expected. \\

\noindent{\bf Majority-spin channel.} Here is it interesting to look at the band structure of Pd/Fe/Ir(111) surface first when SOI is switched off (Fig. 7). Only one intense unoccupied resonant state is found at ${\overline\Gamma}$ point around 0.41 eV. This state is attributed mainly to the hybridization of $p_z$ and $d_{z^2}$ orbitals of Fe and Pd, and therefore has the right symmetry to survive across the vacuum (see Fig. 7(d)). 

By switching on SOI (Fig. 3(e-g), all the bands get pushed to higher energies and one notices the appearance of other surface and resonant states at the $\overline{{\Gamma}}$ point. We distinguish two surface states located within the band gap of the surface-projected band structure with energies 1.27 eV and 1.45 eV. Similarly to what has been predicted in half metals~\cite{Mavropoulos2}, these two states originate from the minority-spin channels (see previous discussion) and are projected via SOI into the majority-spin channel. As mentioned above they originate from the hybridization of $d_{xz}$ and $d_{yz}$ orbitals of Ir and Fe.

The SOI-driven projection mechanism can be grasped by recalling the first order effect of SOI on the electronic structure. The off-diagonal elements of spin-orbit potential in the spinor basis 
\begin{equation}
    V_{\mathrm{SOI}}=\zeta \vec{L}\cdot\vec{\sigma}=\left( {\begin{array}{cc}
   V_{\mathrm{SOI}}^{\uparrow\uparrow} & V_{\mathrm{SOI}}^{\uparrow\downarrow} \\
   V_{\mathrm{SOI}}^{\downarrow\uparrow} & V_{\mathrm{SOI}}^{\downarrow\downarrow} \\
  \end{array} } \right)
\end{equation}
are responsible for flipping the spin. Here $\zeta$ is the SOI strength, $\vec{L}$ and $\vec{\sigma}$ are respectively the angular momentum operator and the Pauli spin matrix. The Hamiltonian without SOI, $H^0$, is spin-diagonal and the corresponding Bloch eigenfunctions are $(\Psi_{n\vec{k}\uparrow}^{(0)},0)$ and $(0,\Psi_{n\vec{k}\downarrow}^{(0)})$, where $n$ is the band index. The Schr\"odinger equation for the perturbed wave function reads

\begin{equation}
 \left(
  {\begin{array}{cc}
   H^{0\uparrow} + V_{\mathrm{SOI}}^{\uparrow\uparrow} - E & V_{\mathrm{SOI}}^{\uparrow\downarrow} \\
   V_{\mathrm{SOI}}^{\downarrow\uparrow} & H^{0\downarrow} +V_{\mathrm{SOI}}^{\downarrow\downarrow} - E \\
  \end{array} } \right)  
  \left(
  {\begin{array}{c}
  \Psi_{n\vec{k}\uparrow} \\
  \Psi_{n\vec{k}\downarrow}
  \end{array} } \right)  = 0.
\end{equation}
Even in an energy window initially without majority-spin states, a component of the spinor gets finite after solving the previous equation in first-order (therefore the index $(1)$):
\begin{equation}
\Psi_{n\vec{k}\uparrow\downarrow}^{(1)} = \sum_n   
\frac{\bra{\Psi_{n\vec{k}\uparrow}^{(0)}}V_{\mathrm{SOI}}^{\uparrow\downarrow}\ket{\Psi_{n\vec{k}\downarrow}^{(0)}}}{E_{n\vec{k}\uparrow}^{(0)} - E_{n\vec{k}\downarrow}^{(0)} }\Psi_{n\vec{k}\downarrow}.
\end{equation}
Thus a weak image of the band structure of the minority-spin channel can be realized in the majority-spin channel (and vice-versa). A quadratic dependence with respect to SOI is expected since the LDOS is related to $|\Psi|^2$. As can be seen from the denominator of Eq.3, the weight of the states will be enhanced close to those k-points where the unperturbed spin-dependent bands cross.  

In addition to the aforementioned surface states we note the presence of resonant states at 0.5 eV, 0.68 eV, which were absent without SOI. Like the states at 1.27 and 1.45 eV, they are probably induced by a projection of the minority-spin states into the majority-spin channel via SOI. The state initally located at 0.41 eV without SOI shifts slightly to a lower energy (0.37 eV). Interestingly these three resonant states (\textit{i.e.} located at 0.37, 0.5 and 0.68 eV) are very much different from those seen in the same energy window of the majority-spin channel of bare Fe/Ir(111) substrate (see Fig. 6(c)), which indicates the large impact of Pd.

\subsection{Ir(111) surface and Pd$_{\mathrm{nMLs}}$/Fe/Ir(111) with $\mathrm{n} =  2,3,5,7$}

To complete our analysis, we increased the number of Pd layers deposited on the substrate. With this, we explore the stability of the features observed with a single Pd overlayer. Also, this allows to better identify the electronic states that belong to Pd and neither to Fe nor to Ir. We note that Fe/Ir(111) covered with Pd films thicker then the so-far discussed single Pd ML can host magnetic skyrmions as already demonstrated for Pd$_{\mathrm{2MLs}}$/Fe/Ir(111)\cite{Crum,Hanneken,Dupe}. When increasing the thickness of Pd films, the XMR signal detecting the presence of magnetic skyrmions will then be mostly coming from the Pd layers. 

Interestingly, the surface-projected band structure of Pd films plotted in Fig. 8 bears several similarities to that of the single Pd-overlayer. For instance, the minority-spin interface resonant state originating from hybridization of the electronic states of Pd and Fe at the ${\overline\Gamma}$ point located at 0.68 eV in the single Pd overlayer seems to shift to a higher energy, around 1 eV and stays there independently from the thickness of Pd. Contrary to the former, the latter state is insensitive to the presence of SOI and becomes a surface state that lies in the band gap of the surface-projected band structure when the number of Pd layers on Fe/Ir(111) substrate is greater than 2. Furthermore, we notice that the position of this state is very similar to the one of the surface state of Pd(111) surface located at 1.3  eV as measured by angle-resolved photoemission~\cite{Hulbert}. 
Moreover, we note that the states found at 1.27 eV and 1.45 eV in Pd/Fe/Ir(111) survive at the IrFePd interface even for thicker Pd films but, and as expected, their intensity on the Pd surface decreases strongly. 

\begin{figure}
\begin{center}
\includegraphics*[angle=0,width=1\linewidth]{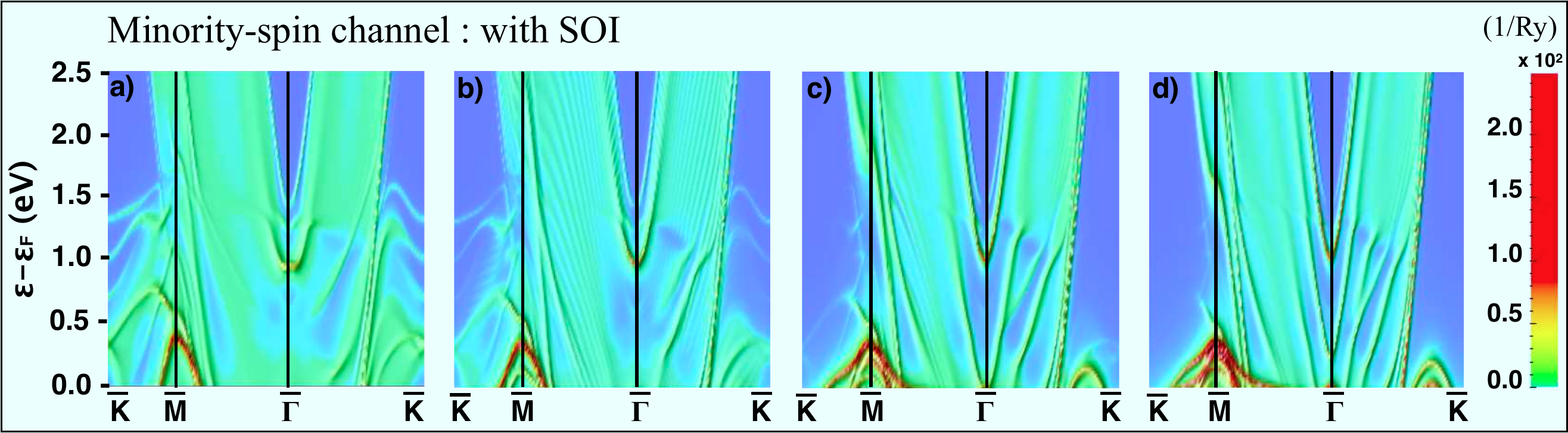}
\end{center}
\caption{Surface-projected band structure in the Pd surface layer for Minority-spin channel in the case of the followig structures: a) Pd$_{\mathrm{2MLs}}$/Fe/Ir(111), b) Pd$_{\mathrm{3MLs}}$/Fe/Ir(111), c) Pd$_{\mathrm{5MLs}}$/Fe/Ir(111) and d) Pd$_{\mathrm{7MLs}}$/Fe/Ir(111).} 
\label{1A1fig8}
\end{figure}

\section*{Summary}
We have presented a detailed study of the electronic structures of thin films of Pd deposited on Fe/Ir(111) surface and focused on the unoccupied states close to the $\overline{\Gamma}$-point of the surface-projected band structure. We found several states having a resonant nature being either confined at PdFe or FeIr interfaces with the possibility of having surface states. These states are located either in the minority-spin or in the majority-spin channel. We investigated their origin and characterized their orbital nature, which dictate the tunneling matrix elements requested in transport experiments based on STM and, thus, are important for the recently discovered XMR-effect. 

We found that the spin-orbit coupling of the Ir substrate not only splits some of these states but also shift them while modifying the curvature of the bands, which affect the electron effective mass and therefore the decay of the states in the vacuum. Interestingly the spin-orbit interaction projects some of the resonant states of the minority-spin channel to the energy-band gap of the majority-spin channel, promoting them then to interface states. 

 \section*{Acknowledgments}
We thank G. Bihlmayer for fruitful discussions.  This work was funded by the European Research Council (ERC) under the European Union's  Horizon 2020 research and innovation programme (ERC-consolidator Grant No. 681405 DYNASORE). We gratefully acknowledge the computing time granted by the JARA-HPC Vergabegremium and VSR commission on the supercomputer JURECA at Forschungszentrum J\"ulich.

\end{document}